\documentclass[a4paper,11pt]{article}
\pdfoutput=1 

\usepackage{jcappub} 

\usepackage[T1]{fontenc} 
\usepackage{graphicx}
\usepackage{float}

\title{\boldmath A Search for AGN sources of the IceCube Diffuse Neutrino Flux}

\author[a]{K. McDonough,}
\author[b,1]{K. Hughes,\note{Corresponding author.}}
\author[a]{D. Smith,}
\author[a]{and A. G. Vieregg}

\affiliation[a]{Department of Physics, Astronomy and Astrophysics, Enrico Fermi Institute, Kavli Institute for Cosmological Physics, University of Chicago, Chicago, IL 60637}
\affiliation[b]{Department of Physics, Pennsylvania State University, University Park, PA, USA}

\emailAdd{kmcd23@uchicago.edu}
\emailAdd{hughes.525@osu.edu}

\abstract{The origin of the diffuse astrophysical neutrino flux measured by the IceCube Observatory remains largely unknown.
Although NGC 1068 and TXS 0506+056 have been identified as potential neutrino sources, the diffuse flux of neutrinos must have additional sources that have not yet been identified.
Here we investigate potential correlations between IceCube's neutrino events and the Fermi and MOJAVE source catalogs, using the publicly-available IceCube data set. 
We perform three separate spatially-dependent, energy-dependent, and time-dependent searches, and find no statistically significant sources outside of NGC 1068. We find that, under the most optimistic assumptions of a spectral index of 2.0 and a neutrino flux uncorrelated with the gamma ray flux, no more than 13\% of IceCube's neutrino flux originates from blazars over the whole sky. Then, using an energy-dependent likelihood analysis, the limit on neutrinos originating from blazars reduces to 9\% in the Northern hemisphere under the same spectral index and flux assumptions. Finally, we set limits on individual sources from the MOJAVE radio catalog after finding no statistically significant time-flaring sources.}

\begin{document}
\maketitle
\flushbottom

\section{Introduction}
The spectrum of IceCube's high-energy neutrino flux, first detected in 2013~\cite{IceCube2013diffuseflux}, is consistent with a power-law extending from tens of TeV to a few PeV, and with flavor ratios consistent with a pion decay origin~\cite{IceCube2016characterists,piondecayicecube, Abbasi_2022}. While searches for point sources contributing to this neutrino flux have largely produced null results, there have been significant excesses from two specific sources. Evidence for the first source, TXS 0506+056, includes a neutrino event coincident in time and space with the flaring blazar  detected by Fermi and MAGIC with a significance of 3 $\sigma$ \cite{icecube2018multimessenger,icecube2018neutrino}. More recently, NGC 1068, a nearby non-blazar active galactic nuclei (AGN), was identified as a potential neutrino source with a significance of 4.2 $\sigma$ \cite{icecube2022ngc1068}. However, these sources alone do not explain the majority of the diffuse neutrino flux, indicating that neutrinos in this energy range are produced by a large number of extragalactic sources \cite{Smith_2021}.

There have been various proposals for possible sources of the diffuse neutrino flux. Gamma-ray bursts (GRBs) \cite{gammaray1,gammaray2,gammaray3,gammaray4}, star-forming galaxies \cite{StarFormingGalaxies}, both blazar and non-blazar AGN \cite{AGN_1991,AGN_1997,Blazars2001}, and fast radio bursts~\cite{Pizzuto:2019kG,Aartsen_2018,Fahey_2017} have all been considered. However, various models have been ruled out as the primary source for IceCube's neutrino flux. In particular, there is an observed lack of correlation with blazar AGN, effectively eliminating flaring blazars as the primary source of the diffuse neutrino flux \cite{Aartsen_2017,Abbasi_2023}. Starburst and other star-forming galaxies are also unable to account for the entirety of this signal without exceeding the measured intensity of the isotropic gamma-ray background \cite{Bechtol_2017,Hooper_2019}. 

Our work in this paper consists of three analyses conducted using the publicly available data set from IceCube, containing 10 years of muon track neutrino events~\cite{icecube10year}. First, we present an update of an earlier analysis that used IceCube's three-year public data set \cite{Smith_2021} to investigate whether a significant fraction of the neutrino flux could originate from blazar or non-blazar AGN. In this updated analysis, we compare IceCube neutrino events from the 10 year catalog to sources in the Fourth Catalog of AGN detected by the Fermi Large Area Telescope (the 4LAC catalog) \cite{4LACcatalog}, and our results are consistent with both the original work and an updated work \cite{Smith_2021,OtherAGN10year}.

We then go on to incorporate neutrino energy information, new in IceCube's 10 year catalog, and perform an energy-dependent analysis with a likelihood formulation that includes the provided ``energy proxy'' of the neutrino, focusing on the 4LAC sources in the Northern sky using the approach outlined in \cite{methodspointsource2008}. We restrict this analysis to the Northern sky, defined conservatively as events with a declination of > 10$^{\circ}$, due to the coarse binning of the smearing matrices provided by the IceCube public data set. These smearing matrices describe the spread in reconstructed energy, direction, and angular uncertainty expected after developing Monte Carlo simulations of the IceCube detector. In the Southern sky, the lack of a dedicated cosmic ray simulation makes characterizing the muon background too challenging, and thus this region is omitted from this analysis. We also choose to omit the horizon region, defined by the smearing matrices as -10$^\circ$ to 10$^\circ$, to be sure any edge effects in the detector response are not impacting our analysis. Thus, only the Northern sky above 10$^\circ$ declination remains in the analysis. Even still, the addition of information about the neutrino energy distribution increases the sensitivity of the search overall; previous work found that with energy information included, the total number of events needed for a 5 $\sigma$ significance is reduced by a factor of two \cite{methodspointsource2008}. The same study also outlines a process that can be repeated in the Southern sky with a more robust energy reconstruction simulation that characterizes the muon background. 

The third analysis presented here is a time-dependent threshold analysis of the MOJAVE XV radio catalog \cite{MOJAVE}, in which, in addition to spatial coincidences, time windows are weighted corresponding to the magnitude of flares coming from a MOJAVE source at that time. A more complete description is included in Section \ref{time_dependent_methods}. The fluctuations seen in the radio emission from these galaxies could be correlated with high-energy neutrino emission; $p-\gamma$ interactions in the sources could create both neutrinos at high energies and photons at lower energies after the initial 
high-energy photon is lost through a chain of pair production~\cite{Plavin_2021}. 

Radio AGN in the MOJAVE radio catalog have previously been investigated as a possible source class for the IceCube neutrino flux in a time-independent analysis \cite{desai2021mojave}, and our addition of a time-dependent likelihood function probes a new signal hypothesis. The multiple epochs present in the MOJAVE data make it an ideal catalog for investigating a possible correlation between neutrinos and radio emission.
We apply the process outlined in \cite{abbasi2021search} for an individual source to the entire MOJAVE catalog.

\section{Method Overview}
\label{section:2.0}
Each analysis presented here uses data from \href{http://doi.org/DOI:10.21234/sxvs-mt83}{IceCube's public data release} consisting of muon track events from April 2008 to July 2018 \cite{icecube10year}. The data is from the 40 string detector in 2008, the 59 string detector in 2009, the 79 string detector in 2010, and the completed 86 string detector for the remaining 7 years. The entire data set contains 1,134,450 muon track events. Each event has a reported direction, angular resolution, time of event, and ``energy proxy,'' which is related to the energy deposited in the detector. Each year of this data set includes an effective area for the detector as determined by simulation, as a function of declination and neutrino energy. 
To test for evidence of a neutrino signal from an individual point source, we follow the approach outlined in~\cite{likelihoodfunc}. The likelihood that a given source results in $n_s$ events, out of a total $N$ recorded in the detector, is given by:
\begin{equation}
\large\mathcal{L} (n_s) = \prod\limits^{N}_{i} 
\Large\left[ \frac{n_s}{N} S_i (|\Vec{x_s}-\Vec{x_i}|)+\left( 1 - \frac{n_s}{N} \right) B_i (\sin{\delta_i} ) \right],
\label{equation1}
\end{equation}

\noindent where $S_i$ and $B_i$ are probability distribution functions (PDFs), $\Vec{x_s}$ is the direction to the source, $\Vec{x_i}$ is the reported direction of the event, and $\delta_i$ is the declination of the event. The signal PDF consists of individual spatial, energy, and time PDFs:

\begin{equation}
\Large S_i = P^{sig}_{i}(\sigma_i, \Vec{x_i}|\Vec{x_s}) \cdot \epsilon^{sig}_{i}(E_i, \delta_i|\gamma) \cdot T^{sig}_{i}(t_i,\phi_s),
\label{equation2}
\end{equation}

\noindent where $P^{sig}_{i}$, $\epsilon^{sig}_{i}$ and $T^{sig}_{i}$ are the respective spatial, energy, and time components for the signal PDF, $\sigma_i$ is the uncertainty associated with each event, given in terms of angular resolution provided by the data set, $E_i$ is the energy proxy, $\gamma$ is the spectral index (set to either 2 or 2.5 for each analysis), $t_i$ is the time of the event, and $\phi_s$ is the flux of the source as a function of time. The background PDF is defined similarly by:

\begin{equation}    
\Large B_i = P^{bkg}_{i}(\delta_i) \cdot \epsilon^{bkg}_{i}(E_i) \cdot T^{bkg}_{i}(t_i),
\label{equation3}
\end{equation}

\noindent where again $P^{bkg}_{i}$, $\epsilon^{bkg}_{i}$, and $T^{bkg}_{i}$ are the respective spatial, energy, and time components for the background PDF. For the work presented in this paper, the spatial component is included in all three analyses, while the energy and time components are only used in the energy-dependent analysis and the MOJAVE catalog study, respectively. In the following sections, the PDFs used in each analysis will be described.

\subsection{Catalog Descriptions}

We compare the measurements from IceCube to the results from two external catalogs: the Fermi 4LAC catalog \cite{4LACcatalog} and the MOJAVE radio catalog \cite{MOJAVE}. 

The Fermi 4LAC catalog is comprised of active galactic nuclei (AGN) sources detected in the 50 MeV to 1 TeV energy range over an eight year period from August 2008 to August 2016. Of the 2863 total AGNs, 2796 are blazars, 63 are non-blazars, and 4 are unidentified. 
The blazars can be further broken down into 658~flat spectrum radio quasars (FSRQs), 1067 BL Lacs, and 1071 ``blazars of unknown origin.'' We do not consider sources within 3$^{\circ}$ of the poles.

The MOJAVE XV catalog consists of observations in the 15~GHz band of 437~AGN over the course of 20 years, between 1996 and 2016. This list of sources includes 265 quasars, 127 BL Lacs, 27 Radio Galaxies, 5 Seyfert-1 Galaxies, and 13 unidentified objects. During  this time, 5321 measurements were taken of the AGN sources, with most sources visited between 5 and 15 times. These time-dependent measurements give unique insight into whether a source is flaring, either relative to its baseline or compared to similar sources. In this study, we only consider MOJAVE data taken after 2008, to align with the operational window of the IceCube instrument.

\section{A Search for Correlations with Blazar and Non-Blazar AGN Using a Spatially-Dependent Likelihood}
\label{section:spatial}
We first search for correlations with blazar and non-blazar AGN using a spatially-dependent likelihood.  
We define the spatial PDF as:

\begin{equation}
\begin{split}
   & \Large P^{sig} = \frac{1}{2\pi \sigma^{2}_{i}}e^{- \frac{|\Vec{x_s}-\Vec{x_i}|^{2}}{2 \sigma^{2}_i}} \\
    & P^{bkg} = \frac{\mathcal{P}_B (\sin\delta_i)}{2\pi},
\end{split}
\label{equation4}
\end{equation}

\noindent The function $\mathcal{P}_B$ is equal to the fraction of events in the data set averaged across a band of $\pm 6^{\circ}$ in declination, $\delta$, around a given source. We only consider sources with declination between $\pm 87^{\circ}$ due to the limited amount of solid angle near the poles with which to characterize the background PDF. To find the value of $n_s$ for a given source, the likelihood function is maximized with respect to the free parameter $n_s$, as outlined in~\cite{methodspointsource2008}. 

\begin{figure}[t!]
\begin{center}
    \includegraphics[width=0.9\textwidth]{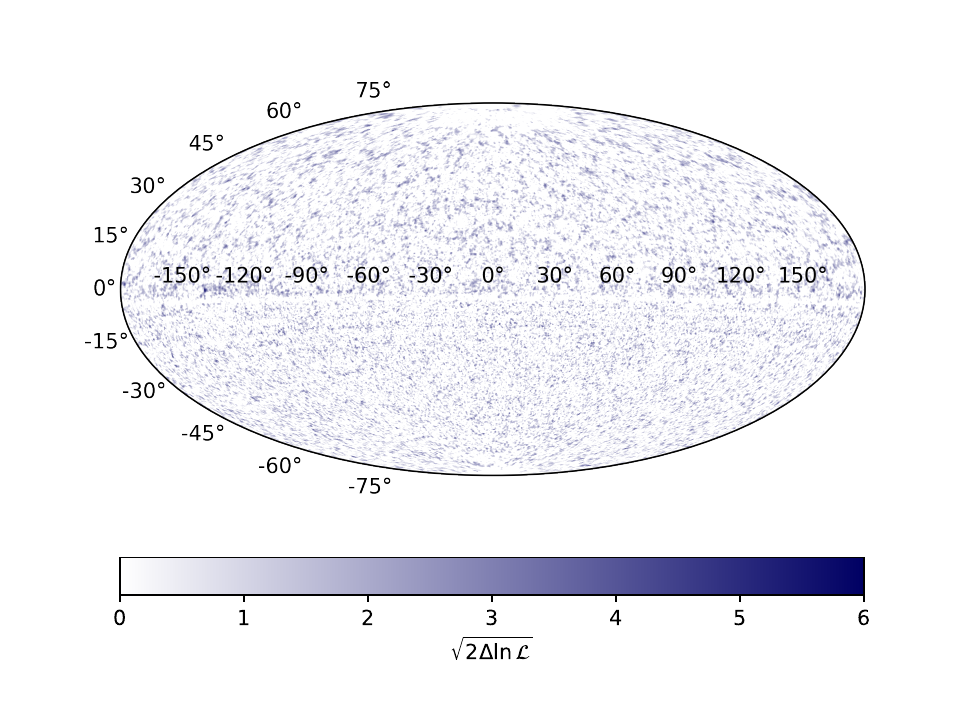}
\vspace*{-10mm}
\caption{An all-sky map of the likelihood distribution of neutrino point sources, $\sqrt{2\Delta\ln\mathcal{L}}$, in Right Ascension (horizontal axis) and Declination (vertical axis) in an Aitoff projection.}
\label{fig:aitoff_spacial}
\end{center}
\end{figure}

The statistical significance of a neutrino point source over a background-only hypothesis is calculated using the following test statistic:

\begin{equation}
    2\Delta\mathcal{L}(n_s) = 2 \left[\ln \mathcal{L}(n_s) - \ln \mathcal{L}(0)\right],
\label{test_statistic}
\end{equation}

\noindent where $\mathcal{L}(0)$ is the likelihood for the background-only hypothesis in which there are no signal events from a given direction. From this, the $p$-value can be calculated by performing an integral over a $\chi^2$ distribution with one degree of freedom. 

Here, the background is being modeled empirically from the experimental data. This is possible due to the dominance of atmospheric muons and has been validated by scrambling the Right Ascension of the dataset and achieving statistically similar maps. The background hypothesis is that there is no correlation between the Fermi 4LAC catalog and the IceCube neutrino dataset; the signal hypothesis is the existence of a statistically-significant correlation between detected neutrino output and gamma-ray emission from known classes of astrophysical objects.

We additionally constrain the fraction of IceCube's diffuse neutrino flux that originates from known classes of astrophysical objects using three different weighting hypotheses for the expected neutrino emission from a given source class. This weighting is applied to $n_s$ in the likelihood, and is determined according to the following hypotheses:
\begin{enumerate}
    \item Gamma-Ray Scaling: The neutrino flux from a given source is proportional to the gamma-ray flux from that source. The gamma-ray flux in the 4LAC catalog is in units of photons between 1-100 GeV per area, per time. This hypothesis would be expected if the gamma-ray emission is primarily produced from hadronic interactions, resulting in a fixed ratio of photons and neutrinos. 
    \item Geometric Scaling: The neutrino flux of a given source is proportional to the inverse of the luminosity distance squared, $1/D_{L}^{2}$. This hypothesis assumes that the neutrino luminosity of a source is only correlated to distance between the source and the detector. This calculation can only be done with sources with measured redshifts, so some sources are excluded in the analysis under this hypothesis.
    \item Flat Scaling: The neutrino flux from a given source is uncorrelated to any other information. 
\end{enumerate}

These hypotheses remain unchanged from our previous two studies \cite{Hooper_2019,Smith_2021}.

\subsection{Results}
\label{spatial_results}
\begin{figure}[ht]
\begin{center}
    \includegraphics[width=0.8\textwidth]{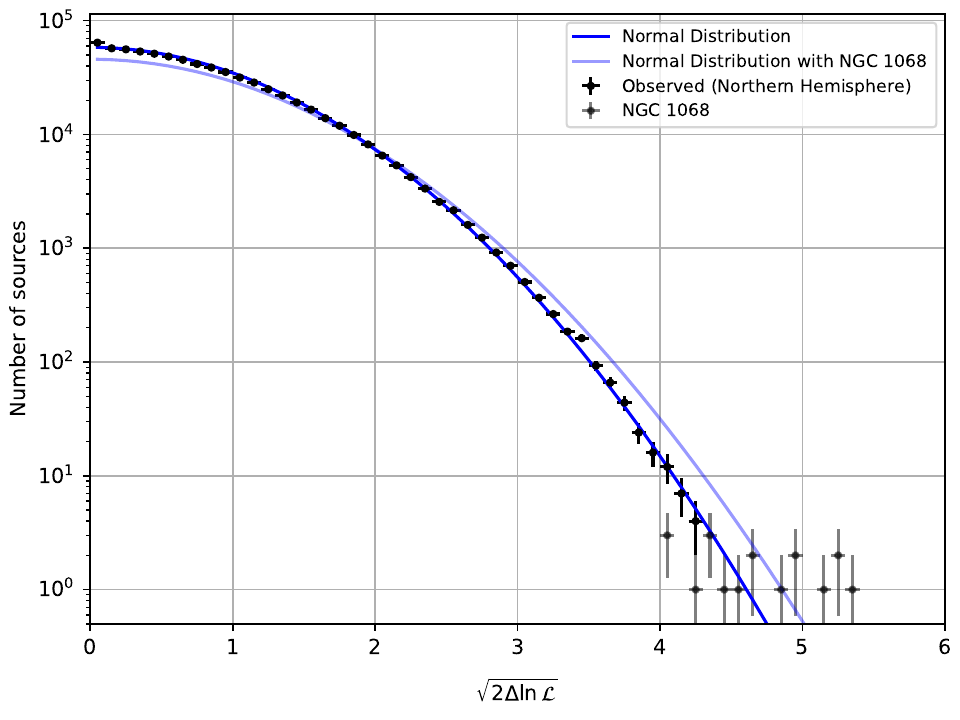}
\caption{The likelihood distribution in favor of a neutrino point source from bins of our all-sky scan. The observed distribution is consistent with a background-only hypothesis when not including NGC 1068 in the best-fit function (dark blue line) and we identify no evidence of another neutrino point source population. Sky locations with $\Delta\ln\mathcal{L}$ < 0, corresponding to a best fit with a negative point source flux, are not shown. The error bars represent the 68\% Poissonian confidence interval on each bin. }
\label{fig:spatial_dist}
\end{center}
\end{figure}

To validate our methodology, in  Fig.~\ref{fig:aitoff_spacial} we show an all-sky map of the likelihood of a neutrino point source, $\sqrt{2\Delta\ln\mathcal{L}}$, in terms of right ascension and declination. The scan was performed in steps of 0.2$^{\circ}$, and at each point we show the value of $n_s$ that maximizes the test statistic defined in Eq. \ref{test_statistic}. In Fig.~\ref{fig:spatial_dist} we show the distribution of this test statistic across the sky. When excluding the points containing NGC 1068, a 4.2~$\sigma$ source discovery by IceCube \cite{icecube2022ngc1068}, the distribution is consistent with a normal distribution. In Table \ref{table:spatial_top_5}, we present the most significant locations in the sky other than NGC~1068, along with the associated likelihoods and pre-trial $p$-values. The post-trial $p$-value for the most significant source is calculated by scrambling the right ascension of the data to create 1000 randomly distributed background-like distributions, and calculating how often these background-like distributions are more extreme than the real data. There is no evidence of a statistically significant source, which is the expected result, consistent with other all-sky scans of the IceCube dataset \cite{IceCube_allsky}.

\begin{table}[h!]
\centering
 \begin{tabular}{|c|c|c|c|c|} 
 \hline
 $2\Delta\ln\mathcal{L}$ & Pre-Trials $p$-value & RA & Dec \\ [0.5ex] 
 \hline\hline
 18.15 & 2.0 $\times$ $10^{-5}$ & 12.60 & -67.40 \\ 
 \hline
 17.72 & 2.6 $\times$ $10^{-5}$ & 296.40 & -21.00 \\
 \hline
 17.55 & 2.8 $\times$ $10^{-5}$ & 300.20 & -32.80\\
 \hline
 17.47 & 2.9 $\times$ $10^{-5}$ & 327.00 & -66.60 \\
 \hline
 17.31 & 3.2 $\times$ $10^{-5}$ & 232.80 & 1.40 \\ [1ex] 
 \hline
 \end{tabular}
 \caption{The five most significant independent locations found in our spatial-only all-sky scan, excluding the coordinates of NGC 1068. The most significant source has a post-trial $p$-value of 0.451, which is not statistically significant.}
 \label{table:spatial_top_5}
\end{table}

\begin{figure}[ht!]
\begin{center}
    \includegraphics[width=\textwidth]{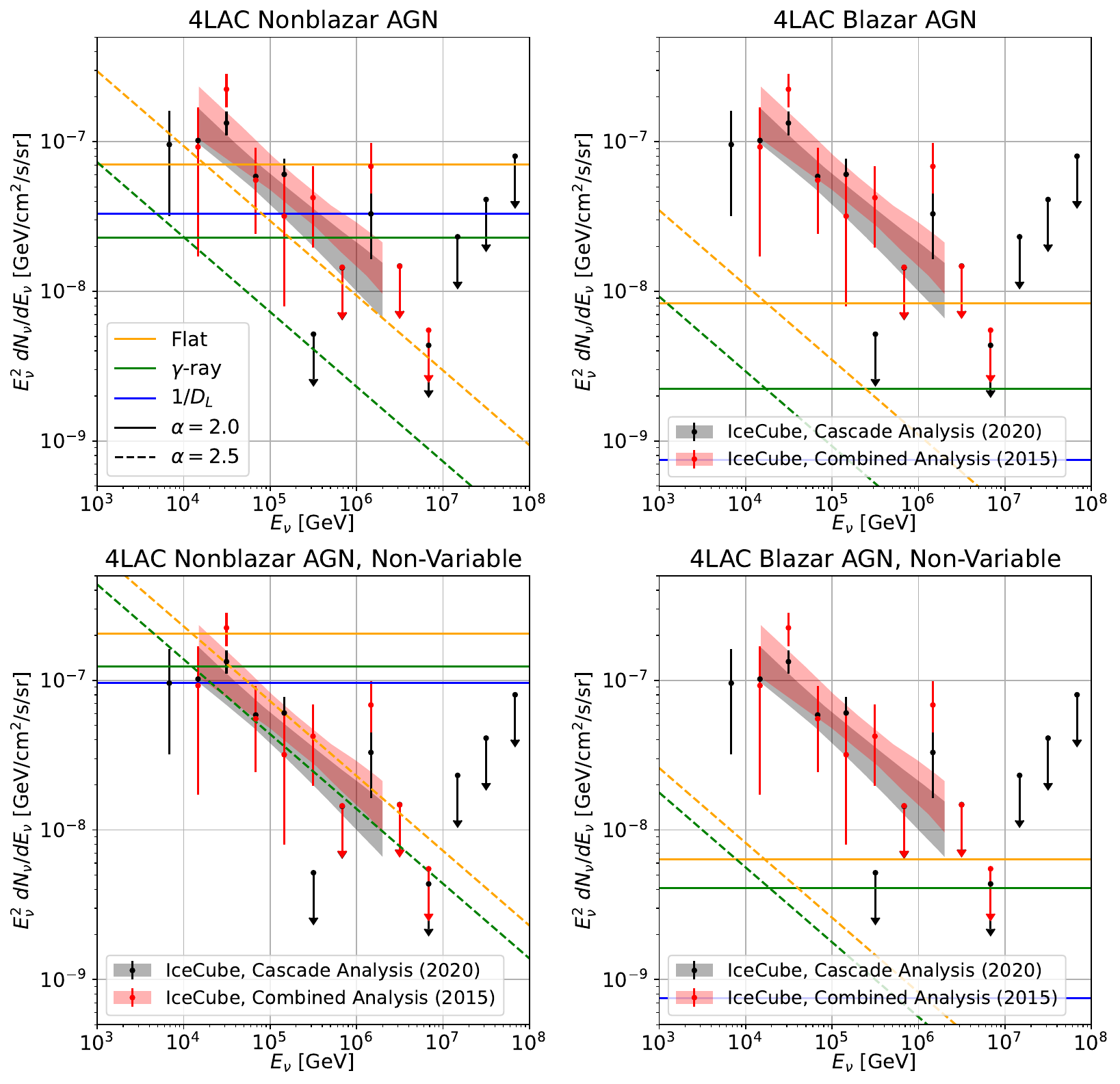}
\caption{The 95\% confidence level upper limits on the neutrino flux for various source classes using the spatial-only likelihood, presented as a function of neutrino energy. We compare these constraints to the diffuse neutrino flux reported by IceCube \cite{icecubediffuse2015,icecubediffuse2020}. In the left frames, we consider non-blazar AGN from the Fermi 4LAC catalog, and in the right frames we consider blazar AGN. In the bottom frames we plot only sources from the catalog that we identify as ``non-variable'' based on the flux density measurements over time, which corresponds to a 4LAC variability index of 18.48. The solid lines are for $\alpha=2.0$, while the dashed lines are for $\alpha=2.5$.  The three different neutrino flux hypotheses (gamma-ray scaling, scaling by luminosity distance, and flat scaling), are shown in green, blue, and yellow respectively. The luminosity distance-scaled upper limits for $\alpha=2.5$ are outside the axes range.}
\label{fig:spatial_4panel}
\end{center}
\end{figure}

We then set limits on the fluxes from specific source classes, calculating the 95\% upper limit as done in \cite{Smith_2021, Hooper_2019}. When considered individually, no source has a statistically significant likelihood after accounting for the appropriate trials factor. When comparing the distances, luminosities, and spectra of the sources with the highest likelihood, we again find no clear features that set these sources apart from other sources in the 4LAC catalog more generally. 

Finding no statistically significant individual sources, we place an upper limit on the neutrino flux from these source classes. In Fig.~\ref{fig:spatial_4panel}, the limits from this analysis on the contribution of blazar and non-blazar AGN to the IceCube diffuse neutrino flux are shown.  We separately show results for sources identified as ``non-variable'' by applying cuts to sources with a variability index below 18.48, as reported by the Fermi Collaboration.  Of the original 2796 blazars and 63 non-blazars, we identify 1674 and 47 ``non-variable'' sources, respectively.  

In all cases, we apply the appropriate completeness factors to account for the incompleteness of the source catalog \cite{Smith_2021}. For blazars, we apply a completeness factor of 1.4. For non-blazars, we apply a completeness factor of 50.6 (154.6) for non-blazar (non-variable) AGN. We place 95\% upper limits (corresponding  $2\Delta\ln\mathcal{L}$ = -3.84) for the three hypotheses outlined in Section \ref{section:spatial}, and for two choices of spectral index. We use a spectral index of $\alpha=2.5$, based on the spectral flux measured by IceCube, and $\alpha=2.0$, the value predicted assuming Fermi acceleration.

From the least constraining upper limits presented, we conclude that blazars can produce no more than 13\% of the diffuse neutrino flux, which is consistent with our previous analysis \cite{Smith_2021} and a publication earlier this year \cite{PhysRevD.103.123018}. This is a slightly more restrictive constraint than our previous result, due to the increased livetime of the 10-year IceCube data. Additionally, we find that under the flat scaling hypothesis, non-variable, non-blazar AGN could produce the entirety of the IceCube diffuse neutrino flux, and that in the case of the other two hypotheses, non-variable, non-blazar AGN could still produce a majority of the flux.

\section{An Energy-Dependent Constraint on the Neutrino Flux from Blazar and Non-Blazar AGN}
\label{sec:energy}
\begin{figure}[ht]
\begin{center}
    \includegraphics[width=\textwidth]{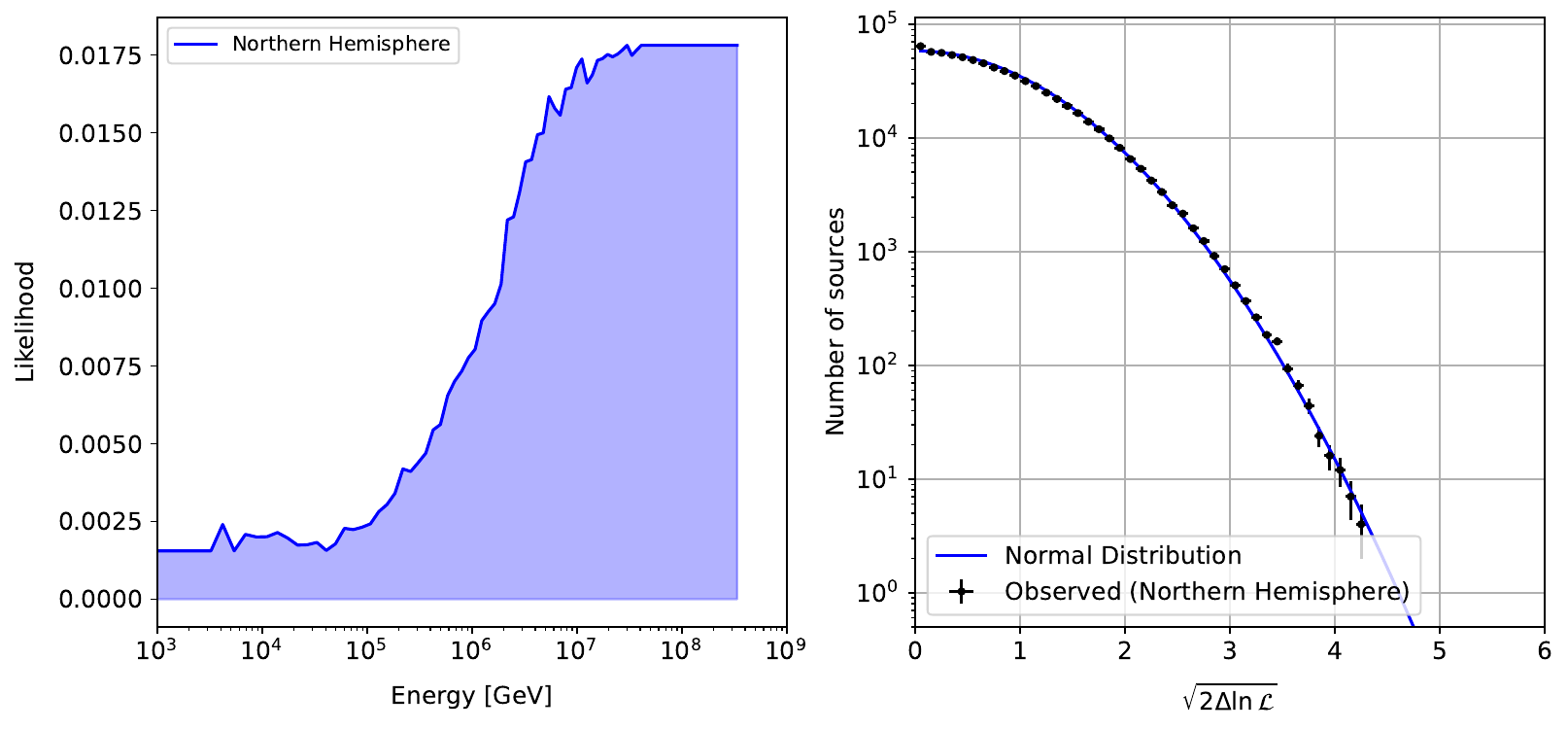}
\vspace*{-5mm}
\caption{Left: The likelihood distribution of the Northern hemisphere data set as a function of the given ``energy proxy'' (in GeV) of the event.  Right: The likelihood distribution in favor of a neutrino point source from the Northern hemisphere all-sky scan in the energy-dependent analysis. The observed distribution is consistent with a background-only hypothesis and we identify no evidence of another neutrino point source population. Sky locations with $\Delta\ln\mathcal{L}$ < 0, corresponding to a best fit with a negative point source flux, are not shown. The error bars represent the 68\% Poissonian confidence interval on each bin.  }
\label{fig:double_likelihood}
\end{center}
\end{figure}

In addition to the spatial-only likelihood analysis described above, we have also perfomed an analysis using an energy-dependent likelihood on data from the Northern hemisphere alone, which unlike the Southern hemisphere is not background-limited by atmospheric muons.
We use the ``energy proxy'' for the muon track events in the IceCube data set, and adopt the smearing matrices from the IceCube data release and a spectral index of $\alpha=2$ to create a function that describes the likelihood of obtaining the reconstructed neutrino energy given the assumed spectral index, as outlined in~\cite{methodspointsource2008}. This signal likelihood is normalized to one and its distribution as a function of energy can be seen in the left panel of Fig.~\ref{fig:double_likelihood}. To speed up the computational process, only events with an energy proxy greater than 100~GeV were included.

Most of the events in this data set are not astrophysical neutrinos; the Northern hemisphere data set is dominated by atmospheric neutrinos, which are a background for this analysis.  Therefore, when constructing the energy-dependent background likelihood distribution, we use the data set itself, dependent on the declination of the point in the sky being tested. 

\subsection{Results}

\begin{figure}[b!]
\begin{center}
    \includegraphics[width=\textwidth]{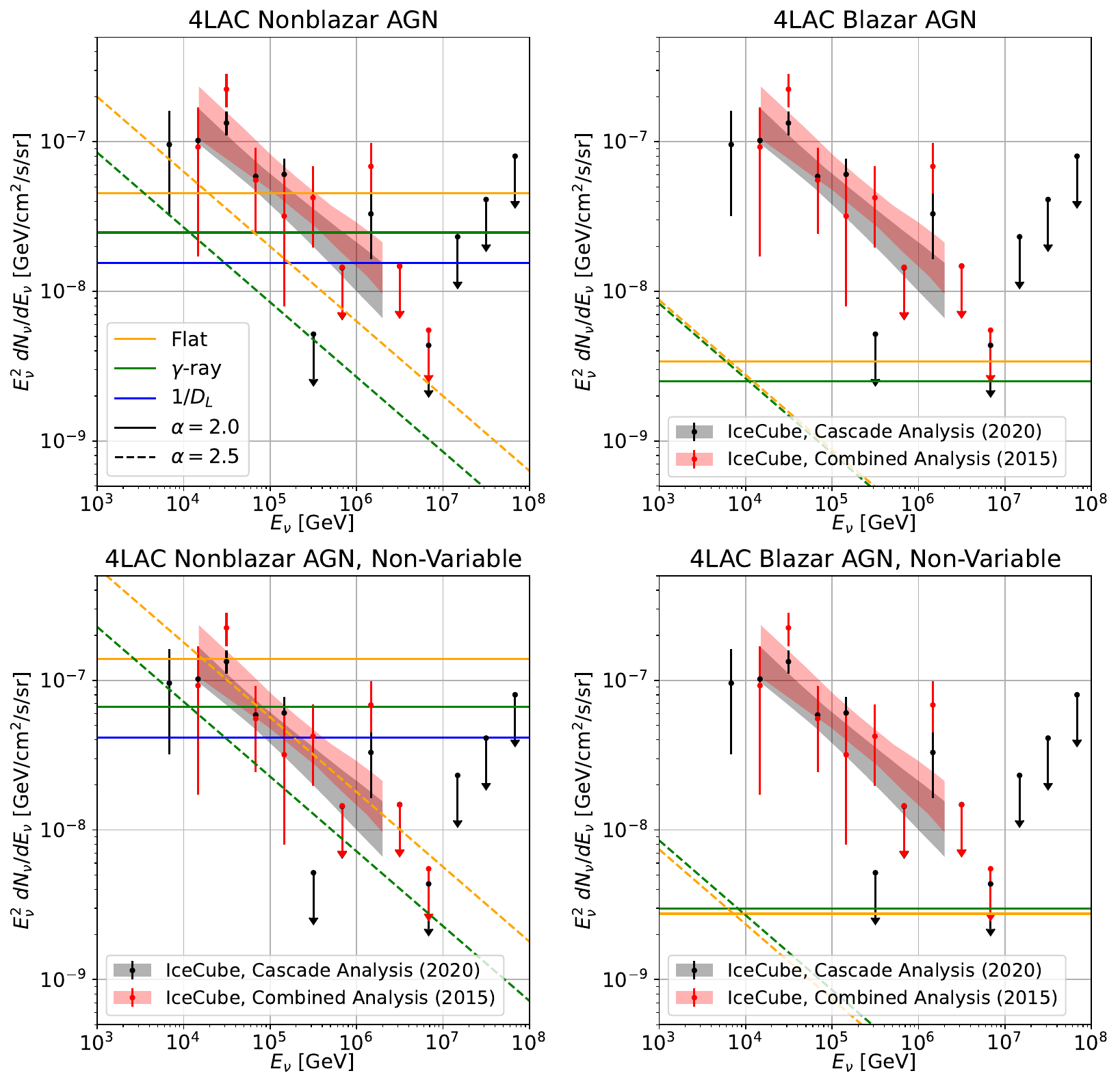}
\caption{The 95\% confidence level upper limits on neutrino flux for various source classes using the spatial and energy-dependent likelihood, presented as a function of neutrino energy. We compare these constraints to the diffuse neutrino flux reported by IceCube \cite{icecubediffuse2015,icecubediffuse2020}. In the left frames, we consider nonblazar AGN from the Fermi 4LAC catalog, and in the right frames we consider blazar AGN. In the bottom frames we plot only sources from the catalog that we identify as ``non-variable'' based on the flux density measurements over time, which corresponds to a 4LAC variability index of~18.48. The solid lines are for $\alpha=2.0$, while the dashed lines are for $\alpha=2.5$.  The three different neutrino flux hypotheses (gamma-ray scaling, scaling by luminosity distance, and flat scaling), are shown in green, blue, and yellow respectively.}
\label{fig:energy_4panel}
\end{center}
\end{figure}

\begin{table}[t!]
\centering
 \begin{tabular}{|c|c|c|c|c|} 
 \hline
 $2\Delta\ln\mathcal{L}$ & Pre-Trials $p$-value & RA & Dec \\ [0.5ex] 
 \hline\hline
 17.47 & 2.9 $\times$ $10^{-5}$ & 0.456 & 182.60 & 39.40\\ 
 \hline
 17.14 & 3.5 $\times$ $10^{-5}$ & 0.458 & 297.20 & 27.60 \\
 \hline
 16.48 & 4.9 $\times$ $10^{-5}$ & 0.459 & 357.20 &15.00\\
 \hline
 16.48 & 4.9 $\times$ $10^{-5}$ & 0.459 & 275.80& 11.40 \\
 \hline
 16.24 & 5.6 $\times$ $10^{-5}$ & 0.461 & 130.80& 56.40\\ [1ex] 
 \hline
 \end{tabular}
 \caption{The five most significant independent locations found in our energy-dependent all-sky scan of only the Northern Hemisphere. The most significant source has a post-trial $p$-value of 0.456, which is not statistically significant.}
 \label{table:energy_top_5}
\end{table}

With the test statistic based on the new likelihood formulation in hand, we find the updated distribution of the test statistic shown in the right panel Fig. \ref{fig:double_likelihood}.
This distribution falls along a normal distribution, consistent with no significant source excesses. The pre- and post-trial $p$-values are calculated in the same was as in Section \ref{spatial_results}. The most significant locations on the sky in this search are presented in Table \ref{table:energy_top_5}. While not statistically significant in this search, we do note that the brightest spot in our likelihood map, with a Right Ascension=182.60 and Declination=39.40, is within two arc minutes of NGC 4151, a source with recent evidence of neutrino emission \cite{neronov2023neutrino}. NGC 1068 falls just below the lower boundary of the analysis in declination and is thus excluded.

We again compare against the Fermi 4LAC catalog with the same source hypotheses outlined in Section~\ref{section:spatial}.
Because we only analyze half the sky, the number of sources in our catalog is decreased by roughly 50\%; however, we also expect the background to decrease by more than 50\%, as the atmospheric events in the Southern hemisphere contribute significantly to the overall background rate at the highest energies. 

The limit set by this energy- and spatial-dependent search in the Northern hemisphere with the 10-year data set are even more restrictive than that of the spatial-only analysis presented in Section~\ref{section:spatial}.
We use the same completeness factor as in the spatial-only analysis with an additional factor of 2 that accounts for only looking in the Northern hemisphere. 

The resulting 95\% upper limits are summarized in Fig. \ref{fig:energy_4panel} for each of the neutrino flux scaling hypotheses.  At most, non-variable blazar AGN could contribute up to 9\% of the neutrino flux. The improved limits for blazar AGN again suggests that the majority of the IceCube neutrino flux is likely originating from a different source class.

The limits set for non-blazar AGN are also slightly improved after folding in the event energy. Even still, non-blazar, non-variable AGN could contribute up to 95\% of the IceCube diffuse neutrino flux under the flat scaling hypothesis. This suggests that non-blazar AGN are still a viable candidate source class.

\section{A Search for Correlations with MOJAVE Radio AGN}
\label{time_dependent_methods}

\begin{figure}[t!]
\begin{center}
    \includegraphics[width=0.8\textwidth]{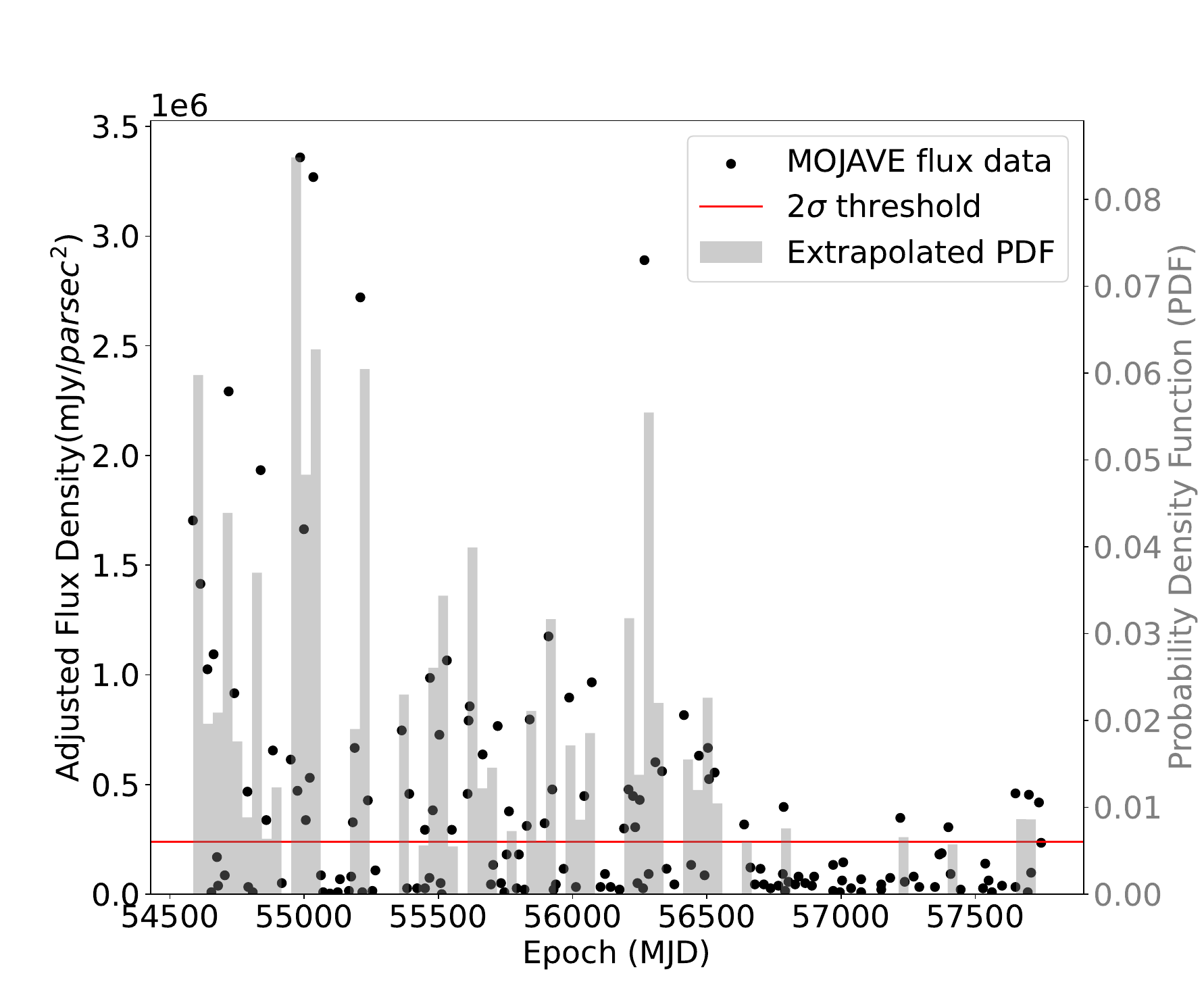}
\caption{The flux density, adjusted for luminosity distance, across the MOJAVE AGN catalog as a function of Modified Julian Dat (MJD). Also shown is the extrapolated PDF. For time periods with no flaring sources, defined as $2\sigma$ above the average recorded flux, the PDF is equal to 0. }
\label{fig:MOJAVE_flux}
\end{center}
\end{figure}

We apply a triggered time-dependent analysis on the MOJAVE catalog, which contains measurements of the radio emission from various AGN over a period of two decades, including one decade of overlap with the IceCube observatory. The time-dependent component of the likelihood $T^{sig}$ is created by calculating the mean value of the flux density for all sources in the MOJAVE catalog after being scaled appropriately with respect to their luminosity distance from Earth. Then, sources are identified as flaring if their flux density at a given time is greater than 2~$\sigma$ above the average flux density. This is illustrated in Figure \ref{fig:MOJAVE_flux}.

We perform the analysis using 436 radio-loud AGN from the MOJAVE catalog. Due to detector location and limits, the catalog only consists of sources higher than -30$^{\circ}$ in declination. Each source in the catalog has a minimum of 5 flux density recordings at different times.
Two sources are eliminated from the analysis, MOJAVE source 0415+379 and source 2200+420, as all of their flux density recordings were above the 2 $\sigma$ threshold and can thus be categorized as always flaring. These two sources could be analyzed independently of this temporal analysis. This methodology was chosen to fairly assess all of the sources in the MOJAVE catalog, which can individually have as few as five data points and as many as 30 over the two decades of data.

\begin{figure}[ht]
\begin{center}
    \includegraphics[width=0.7\textwidth]{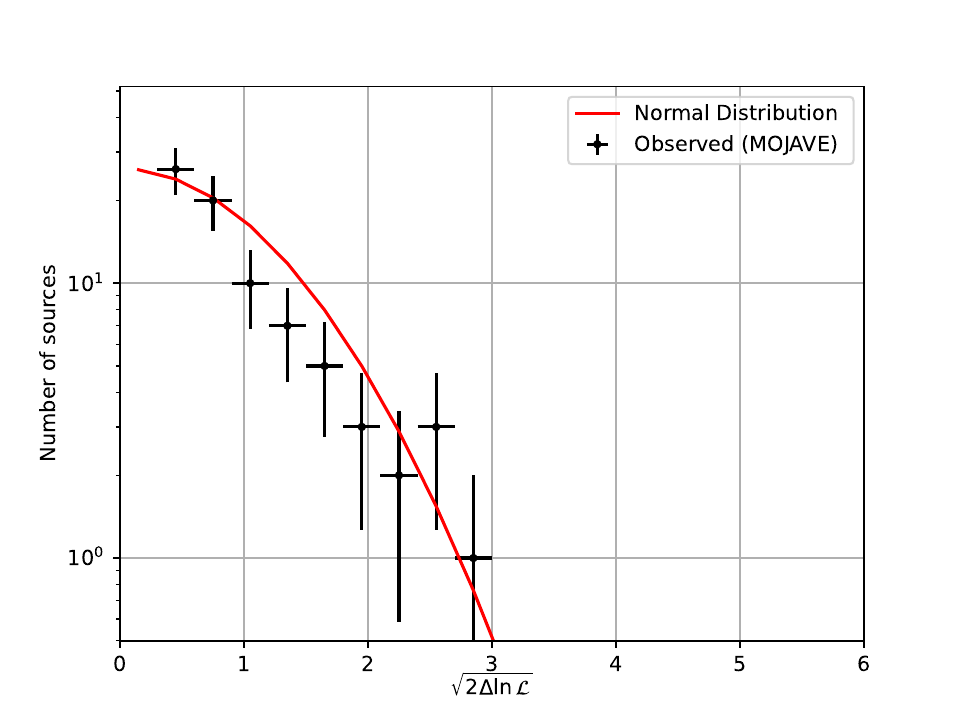}
\caption{The likelihood distribution in favor of a neutrino point source at the locations of the MOJAVE sources. The observed distribution is consistent with a background-only hypothesis, with a slight dip in the 1-2 $\sqrt{2\Delta\ln\mathcal{L}}$ range, and we identify no evidence of another neutrino point source population. Sky locations with $\Delta\ln\mathcal{L}$ < 0, corresponding to a best fit with a negative point source flux, are not shown. The error bars represent the 68\% Poissonian confidence interval on each bin. }
\label{mojave_dist}
\end{center}
\end{figure}

Once the threshold on the catalog is set, any flux density point measured under the threshold is given a time likelihood value of 0. Above the threshold, we bin the data into 0.1~year bins, as outlined in \cite{Plavin_2021} as the smallest reasonable bin size for Radio AGN. This was selected to ensure that there is minimal likelihood overlap, as larger time bins could result in neutrinos and flaring sources being erroneously correlated.
After binning the flux density, the total flux density in all the bins is normalized to one. This creates the time-dependent portion of the signal PDF, where the likelihood for each muon track neutrino event is assigned based on the trigger time of the IceCube event. This likelihood is used in combination with the spatially-dependent likelihood described in Section \ref{section:spatial}.  The delay between the arrival of the neutrino and the photons from the source is on the scale of seconds and can be ignored, as it is absorbed in the 0.1 year binning. 


The time-dependent component of the background PDF is independent of seasonal variations \cite{abbasi2021search,abbasi2011time}, but is dependent on the declination and year of the neutrino event. An additional time-dependence arises due to the construction of IceCube itself: more atmospheric muons were recorded in the first three years of this dataset than later years due to the phased construction of the detector. This prevalence of events with declination <0$^{\circ}$ during these early years is described in the background PDF, with unique PDFs defined as a function of declination for each 0.1 year bin. After the first 3 years, the PDF no longer changes with time, only declination.

\subsection{Results}
In Fig. \ref{mojave_dist}, we show the likelihood distribution of the MOJAVE sources tested.  After accounting for trials factors, no statistically significant sources are identified.

\begin{figure}[ht]
\begin{center}
    \makebox[\textwidth][c]{\includegraphics[width=\textwidth]{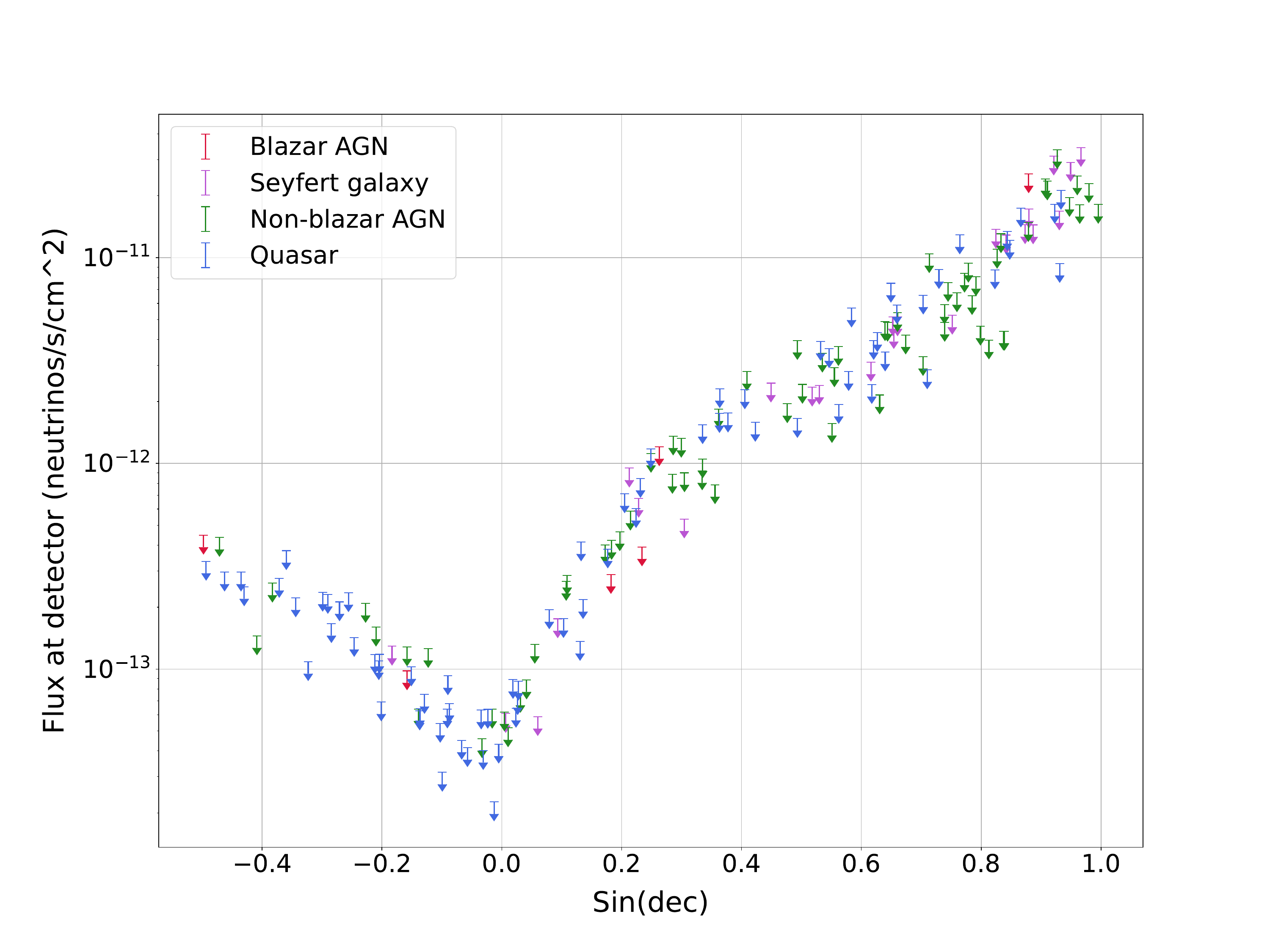}}%
\caption{The 95\% confidence level upper limits on the neutrino flux coming from the 175 MOJAVE sources that had an $n_s$ > 0  and $\sqrt{2\delta\ln\mathcal{L}}$ > 0, as a function of declination. The shape of the limits match the declination-dependent effective volume of the IceCube instrument. The entirety of this list with the flux upper limits, reference name, and source class is in Appendix~\ref{sec:appendix}.}
\label{fig:upperlims}
\end{center}
\end{figure}

We also consider each source individually to set independent limits on the expected flux of neutrinos from each source. These limits are shown in Fig.~\ref{fig:upperlims}.
The list of these sources with their reference name, coordinates, and upper limit neutrino flux density can be found in Appendix~\ref{sec:appendix}. Because we only consider individual sources, rather than source classes, no completeness factor is required.
The MOJAVE catalog includes sources in four different subclasses: quasars, non-blazar AGN, blazar AGN, and Seyfert galaxies. Across each of these subclasses, there is no obvious relationship or trend between source type and neutrino flux limit. 

This time-dependent method could be further refined in future work by varying the time window used to define a flaring source or by combining it with the energy-dependent method discussed in Section \ref{sec:energy}.

\section{Conclusion}
The analyses presented here utilize spatial, energy-dependent, and time-dependent likelihood functions to compare the IceCube diffuse neutrino flux to the Fermi 4LAC and MOJAVE catalogs. We confirm that blazar AGN are unlikely to explain the entirety of the IceCube neutrino flux, while non-blazar AGN are not ruled out. Adding an energy dependence to the previously described spatial analysis tightened constraints on the flux fraction for all source classes tested. The time-dependent analysis with the MOJAVE catalog sets neutrino flux limits for flaring sources broken down by source class.

While we are unable to definitively confirm a specific origin for the diffuse neutrino flux, we are hopeful that future analyses will utilize increased livetime, more complete source catalogs, and a substantive cosmic ray simulation to improve these results further. Additionally, considering specific theoretical models for source neutrino production as a function of neutrino energy could be included to further constrain possible sources.  The time-dependent analysis framework could be used in tandem with other time-dependent catalogs as they become available.

\newpage
\bibliographystyle{JHEP}
\bibliography{references.bib}

\newpage

\appendix
\section{Appendix: MOJAVE Source Limits}
\label{sec:appendix}
\begin{figure}[p]
\begin{center}
    \makebox[\textwidth][c]{\includegraphics[width=\textwidth]{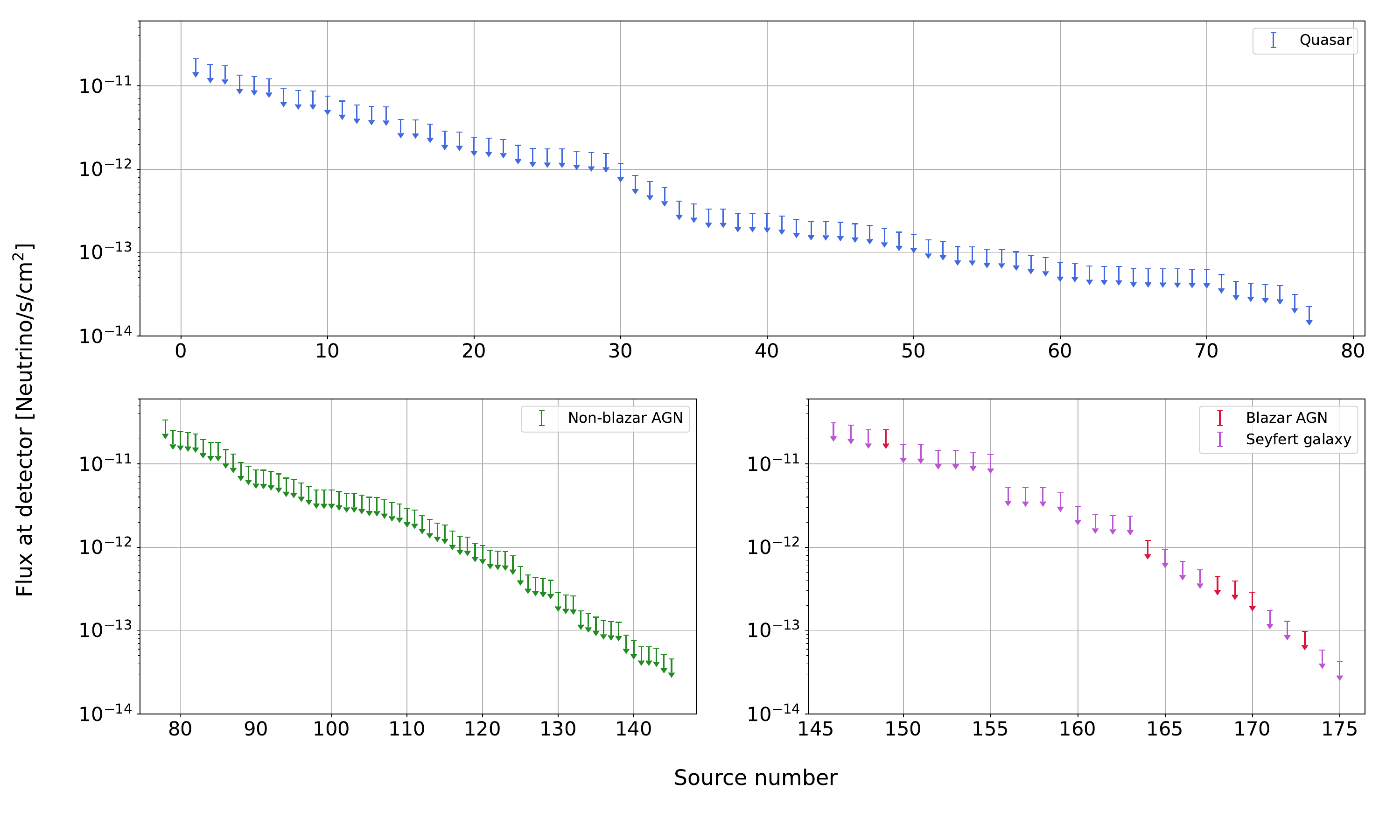}}%
\caption{The 95\% confidence level upper limits on the neutrino flux coming from the 175 MOJAVE sources that had an $n_s$ > 0  and $\sqrt{2\delta\ln\mathcal{L}}$ > 0. On the x axis is the source number and on the y axis is the neutrino flux from the source at the detector. The sources are presented in descending flux order and sorted by source type. The entirety of this list with the flux upper limits, reference name, and source class is in Appendix~\ref{sec:appendix}.}
\label{fig:upperlims_app}
\end{center}
\end{figure}
 
    \centering

    \begin{tabular}{|l|l|l|l|}
        \hline
        Source Number & Flux Upper Limit [$s^{-1}cm^{-2}$] & MOJAVE Reference & Source Class \\ \hline
        1 & 2.12e-11 & 1145-071 & QSO \\ 
        2 & 1.82e-11 & 0106+013 & QSO \\ 
        3 & 1.75e-11 & 1329-049 & QSO \\ 
        4 & 1.34e-11 & 1149-084 & QSO \\ 
        5 & 1.29e-11 & 0414-189 & QSO \\ 
        6 & 1.21e-11 & 0203-120 & QSO \\ 
        7 & 9.37e-12 & 0847-120 & QSO \\ 
        8 & 8.75e-12 & 0752-116 & QSO \\ 
        9 & 8.72e-12 & 1236+077 & QSO \\ 
        10 & 7.51e-12 & 0859-140 & QSO \\ 
        11 & 6.58e-12 & 2345-167 & QSO \\ 
        12 & 5.9e-12 & 0027+056 & QSO \\ 
        13 & 5.69e-12 & 1047+048 & QSO \\ 
        14 & 5.61e-12 & 0607-157 & QSO \\ 
        15 & 3.96e-12 & 1908-201 & QSO \\ 
        16 & 3.92e-12 & 1504-166 & QSO \\ 
        17 & 3.48e-12 & 1127-145 & QSO \\ 
        18 & 2.85e-12 & 1341-171 & QSO \\ 
        19 & 2.8e-12 & 1622-253 & QSO \\ 
        20 & 2.42e-12 & 2328-220 & QSO \\ 
        21 & 2.36e-12 & 1920-211 & QSO \\ 
        22 & 2.28e-12 & 1244-255 & QSO \\ 
        23 & 1.93e-12 & 0142-278 & QSO \\ 
        24 & 1.78e-12 & 0256+075 & QSO \\ 
        25 & 1.76e-12 & 1034-293 & QSO \\ 
        26 & 1.75e-12 & 0742+103 & QSO \\ 
        27 & 1.65e-12 & 2331+073 & QSO \\ 
        28 & 1.58e-12 & 1551+130 & QSO \\ 
        29 & 1.54e-12 & 0119+115 & QSO \\ 
        30 & 1.18e-12 & 0229+131 & QSO \\ 
        31 & 8.45e-13 & 2136+141 & QSO \\ 
        32 & 7.11e-13 & 0710+196 & QSO \\ 
        33 & 6.03e-13 & 1441+252 & QSO \\ 
        34 & 4.15e-13 & 2113+293 & QSO \\ 
        35 & 3.84e-13 & 2223+210 & QSO \\ 
        36 & 3.34e-13 & 1324+224 & QSO \\ 
        37 & 3.33e-13 & 1049+215 & QSO \\ 
        38 & 2.96e-13 & 1611+343 & QSO \\ 
        39 & 2.96e-13 & 2209+236 & QSO \\ 
        40 & 2.92e-13 & 0716+332 & QSO \\ 
        41 & 2.76e-13 & 1633+382 & QSO \\ 
        42 & 2.51e-13 & 0109+351 & QSO \\ 
        43 & 2.36e-13 & 0650+453 & QSO \\ 
        44 & 2.35e-13 & 1638+398 & QSO \\ 

    \end{tabular}


    \centering

    \begin{tabular}{|l|l|l|l|}
        Source Number & Flux Upper Limit [$s^{-1}cm^{-2}$] & MOJAVE Reference & Source Class \\ \hline
        45 & 2.31e-13 & 0110+318 & QSO \\  
        46 & 2.22e-13 & 1417+385 & QSO \\ 
        47 & 2.12e-13 & 1758+388 & QSO \\ 
        48 & 1.94e-13 & 1015+359 & QSO \\ 
        49 & 1.76e-13 & 1030+415 & QSO \\ 
        50 & 1.66e-13 & 0917+449 & QSO \\ 
        51 & 1.42e-13 & 2005+403 & QSO \\ 
        52 & 1.36e-13 & 0954+556 & QSO \\ 
        53 & 1.18e-13 & 0859+470 & QSO \\ 
        54 & 1.18e-13 & 1716+686 & QSO \\ 
        55 & 1.1e-13 & 0850+581 & QSO \\ 
        56 & 1.09e-13 & 0804+499 & QSO \\ 
        57 & 1.02e-13 & 1602+576 & QSO \\ 
        58 & 9.27e-14 & 0646+600 & QSO \\ 
        59 & 8.73e-14 & 0224+671 & QSO \\ 
        60 & 7.52e-14 & 1642+690 & QSO \\ 
        61 & 7.43e-14 & 1532+016 & QSO \\ 
        62 & 6.92e-14 & 0113-118 & QSO \\ 
        63 & 6.85e-14 & 0837+012 & QSO \\ 
        64 & 6.8e-14 & 1128-047 & QSO \\ 
        65 & 6.44e-14 & 0906+015 & QSO \\ 
        66 & 6.43e-14 & 0805-077 & QSO \\ 
        67 & 6.4e-14 & 0015-054 & QSO \\ 
        68 & 6.37e-14 & 0420-014 & QSO \\ 
        69 & 6.32e-14 & 2258-022 & QSO \\ 
        70 & 6.26e-14 & 1406-076 & QSO \\ 
        71 & 5.45e-14 & 1118-056 & QSO \\ 
        72 & 4.51e-14 & 1741-038 & QSO \\ 
        73 & 4.31e-14 & 0440-003 & QSO \\ 
        74 & 4.15e-14 & 2320-035 & QSO \\ 
        75 & 4.02e-14 & 0336-019 & QSO \\ 
        76 & 3.15e-14 & 0539-057 & QSO \\ 
        77 & 2.26e-14 & 0743-006 & QSO \\ 
        78 & 3.35e-11 & 0636+680 & Non-Blazar AGN \\ 
        79 & 2.5e-11 & 0212+735 & Non-Blazar AGN \\ 
        80 & 2.42e-11 & 1959+650 & Non-Blazar AGN \\ 
        81 & 2.36e-11 & 0954+658 & Non-Blazar AGN \\ 
        82 & 2.29e-11 & 1803+784 & Non-Blazar AGN \\ 
        83 & 1.96e-11 & 0238+711 & Non-Blazar AGN \\ 
        84 & 1.82e-11 & 0454+844 & Non-Blazar AGN \\ 
        85 & 1.81e-11 & 1027+749 & Non-Blazar AGN \\ 
        86 & 1.48e-11 & 1542+616 & Non-Blazar AGN \\ 
        87 & 1.31e-11 & 1557+565 & Non-Blazar AGN \\ 
        88 & 1.05e-11 & 1726+455 & Non-Blazar AGN \\ 
        89 & 9.4e-12 & 0846+513 & Non-Blazar AGN \\  
    \end{tabular}


    \centering

    \begin{tabular}{|l|l|l|l|}
        Source Number & Flux Upper Limit [$s^{-1}cm^{-2}$] & MOJAVE Reference & Source Class \\ \hline
        90 & 8.49e-12 & 0128+554 & Non-Blazar AGN \\
        91 & 8.41e-12 & 0708+506 & Non-Blazar AGN \\ 
        92 & 8.08e-12 & 0806+524 & Non-Blazar AGN \\ 
        93 & 7.58e-12 & 1656+482 & Non-Blazar AGN \\ 
        94 & 6.75e-12 & 1011+496 & Non-Blazar AGN \\ 
        95 & 6.55e-12 & 2344+514 & Non-Blazar AGN \\ 
        96 & 5.9e-12 & 1738+476 & Non-Blazar AGN \\ 
        97 & 5.41e-12 & 1206+416 & Non-Blazar AGN \\ 
        98 & 4.89e-12 & 1652+398 & Non-Blazar AGN \\ 
        99 & 4.86e-12 & 1722+401 & Non-Blazar AGN \\ 
        100 & 4.85e-12 & 0603+476 & Non-Blazar AGN \\ 
        101 & 4.65e-12 & 1250+532 & Non-Blazar AGN \\ 
        102 & 4.39e-12 & 1823+568 & Non-Blazar AGN \\ 
        103 & 4.39e-12 & 0613+570 & Non-Blazar AGN \\ 
        104 & 4.21e-12 & 0814+425 & Non-Blazar AGN \\ 
        105 & 3.98e-12 & 1418+546 & Non-Blazar AGN \\ 
        106 & 3.96e-12 & 0912+297 & Non-Blazar AGN \\ 
        107 & 3.7e-12 & 0321+340 & Non-Blazar AGN \\ 
        108 & 3.43e-12 & 1308+326 & Non-Blazar AGN \\ 
        109 & 3.31e-12 & 0621+446 & Non-Blazar AGN \\ 
        110 & 2.92e-12 & 2023+335 & Non-Blazar AGN \\ 
        111 & 2.79e-12 & 1040+244 & Non-Blazar AGN \\ 
        112 & 2.42e-12 & 1215+303 & Non-Blazar AGN \\ 
        113 & 2.15e-12 & 0133+388 & Non-Blazar AGN \\ 
        114 & 1.95e-12 & 1404+286 & Non-Blazar AGN \\ 
        115 & 1.84e-12 & 0518+211 & Non-Blazar AGN \\ 
        116 & 1.56e-12 & 0619+334 & Non-Blazar AGN \\ 
        117 & 1.36e-12 & 0235+164 & Non-Blazar AGN \\ 
        118 & 1.32e-12 & 2201+171 & Non-Blazar AGN \\ 
        119 & 1.12e-12 & 0722+145 & Non-Blazar AGN \\ 
        120 & 1.05e-12 & 1741+196 & Non-Blazar AGN \\ 
        121 & 9.2e-13 & 1514+197 & Non-Blazar AGN \\ 
        122 & 9e-13 & 1717+178 & Non-Blazar AGN \\ 
        123 & 8.84e-13 & 2013+163 & Non-Blazar AGN \\ 
        124 & 7.88e-13 & 0859+210 & Non-Blazar AGN \\ 
        125 & 5.86e-13 & 1228+126 & Non-Blazar AGN \\ 
        126 & 4.65e-13 & 0446+112 & Non-Blazar AGN \\ 
        127 & 4.36e-13 & 2247-283 & Non-Blazar AGN \\ 
        128 & 4.22e-13 & 1940+104 & Non-Blazar AGN \\ 
        129 & 4.02e-13 & 0754+100 & Non-Blazar AGN \\ 
        130 & 2.85e-13 & 1811+062 & Non-Blazar AGN \\ 
        131 & 2.67e-13 & 1038+064 & Non-Blazar AGN \\ 
        132 & 2.62e-13 & 0823-223 & Non-Blazar AGN \\ 
        133 & 1.72e-13 & 0403-132 & Non-Blazar AGN \\ 
        134 & 1.6e-13 & 2128-123 & Non-Blazar AGN \\ 
    \end{tabular}


    \centering

    \begin{tabular}{|l|l|l|l|}
        Source Number & Flux Upper Limit [$s^{-1}cm^{-2}$] & MOJAVE Reference & Source Class \\ \hline 
        135 & 1.45e-13 & 0301-243 & Non-Blazar AGN \\ 
        136 & 1.32e-13 & 0823+033 & Non-Blazar AGN \\ 
        137 & 1.28e-13 & 0903-088 & Non-Blazar AGN \\ 
        138 & 1.26e-13 & 0845-068 & Non-Blazar AGN \\ 
        139 & 8.9e-14 & 0111+021 & Non-Blazar AGN \\ 
        140 & 7.62e-14 & 0808+019 & Non-Blazar AGN \\ 
        141 & 6.39e-14 & 0939-077 & Non-Blazar AGN \\ 
        142 & 6.38e-14 & 0723-008 & Non-Blazar AGN \\ 
        143 & 6.19e-14 & 1514+004 & Non-Blazar AGN \\ 
        144 & 5.18e-14 & 0422+004 & Non-Blazar AGN \\ 
        145 & 4.58e-14 & 2131-021 & Non-Blazar AGN \\ 
        146 & 3.12e-11 & 1849+670 & Seyfert\_1 \\ 
        147 & 2.91e-11 & 1458+718 & Seyfert\_1 \\ 
        148 & 2.57e-11 & 2043+749 & Seyfert\_1 \\ 
        149 & 2.56e-11 & 0106+612 & Blazar \\ 
        150 & 1.73e-11 & 2021+614 & Seyfert\_2 \\ 
        151 & 1.69e-11 & 1700+685 & Seyfert\_1 \\ 
        152 & 1.45e-11 & 1030+611 & Seyfert\_1 \\ 
        153 & 1.44e-11 & 0241+622 & Seyfert\_1 \\ 
        154 & 1.37e-11 & 0831+557 & Seyfert\_2 \\ 
        155 & 1.29e-11 & 1637+574 & Seyfert\_1 \\ 
        156 & 5.25e-12 & 1828+487 & Seyfert\_1 \\ 
        157 & 5.17e-12 & 1957+405 & Seyfert\_2 \\ 
        158 & 5.16e-12 & 0309+411 & Seyfert\_1 \\ 
        159 & 4.47e-12 & 0010+405 & Seyfert\_1 \\ 
        160 & 3.11e-12 & 0415+379 & Seyfert\_1 \\ 
        161 & 2.46e-12 & 1607+268 & Seyfert\_2 \\ 
        162 & 2.4e-12 & 1901+319 & Seyfert\_1 \\ 
        163 & 2.35e-12 & 0738+313 & Seyfert\_1 \\ 
        164 & 1.2e-12 & 0202+149 & Blazar \\ 
        165 & 9.49e-13 & 1345+125 & Seyfert\_2 \\ 
        166 & 6.77e-13 & 0838+133 & Seyfert\_1 \\ 
        167 & 5.37e-13 & 2141+175 & Seyfert\_1 \\ 
        168 & 4.48e-13 & 1622-297 & Blazar \\ 
        169 & 3.93e-13 & 0528+134 & Blazar \\ 
        170 & 2.88e-13 & 1502+106 & Blazar \\ 
        171 & 1.76e-13 & 0430+052 & Seyfert\_1 \\ 
        172 & 1.29e-13 & 1302-102 & Seyfert\_1 \\ 
        173 & 9.81e-14 & 1510-089 & Blazar \\ 
        174 & 5.87e-14 & 1502+036 & Seyfert\_1 \\ 
        175 & 4.23e-14 & 0946+006 & Seyfert\_1 \\ 
    \hline
    \end{tabular}








\end{document}